\documentclass[preprint,draft,prl,10pt,twocolumn]{revtex4}

\input{tcilatex}

\begin{document}

\title{Experimental Realization of a Photonic Bell-State Analyzer}
\author{Philip Walther$^{1}$ and Anton Zeilinger$^{1,2}$}
\affiliation{$^1$Institut f\"ur Experimentalphysik, Universit\"at Wien, Boltzmanngasse 5,
A--1090 Wien, Austria\\
$^{2}$Institut f\"{u}r Quantenoptik und Quanteninformation, \"{O}%
sterreichische Akademie der Wissenschaften, Wien, Austria}

\begin{abstract}
Efficient teleportation is a crucial step for quantum computation and
quantum networking. In the case of qubits, four different entangled Bell
states have to be distinguished. We have realized a probabilistic, but in
principle deterministic, Bell-state analyzer for two photonic quantum bits
by the use of a non-destructive controlled-NOT (CNOT) gate based on entirely
linear optical elements. This gate was capable of distinguishing between all
of the Bell states with higher than 75\% fidelity without any noise
substraction due to utilizing quantum interference effects.
\end{abstract}

\maketitle

Many quantum communication and quantum computation schemes, including
quantum teleportation \cite{bennbrass,zeil}, dense coding \cite%
{densetheory,denseexp}, and quantum cryptography \cite%
{cryptekert,gisin01,jennewein} are based on the maximally-entangled
two-particle quantum states called Bell states. For both the creation and
discrimination of these states, two-qubit quantum logic gates are necessary.
For quantum communication the use of single photons to encode quantum
information is most promising due to the their robustness against
decoherence and the possibility of photon broadcasting. However, it has been
very difficult to achieve the requisite logic operations between two
individual photonic qubits due to their weak physical interaction \cite%
{kim01}.

Remarkably, Knill, Laflamme, and Milburn \cite{knill01} found a way to
circumvent this problem and to implement efficient quantum computation based
on linear optics. They showed that the measurement process itself was enough
to induce strong nonlinearities at the single-photon level.

One of the most important two-qubit gates for quantum computation is the
controlled-NOT (CNOT) gate, which flips the second (target) bit if and only
if the first (control) bit has the logical value $1$, while the control bit
remains unaffected: $\left| 0\right\rangle _{c}\left| 0\right\rangle
_{t}\rightarrow \left| 0\right\rangle _{c}\left| 0\right\rangle _{t}$, $%
\left| 0\right\rangle _{c}\left| 1\right\rangle _{t}\rightarrow \left|
0\right\rangle _{c}\left| 1\right\rangle $, $\left| 1\right\rangle
_{c}\left| 0\right\rangle _{t}\rightarrow \left| 1\right\rangle _{c}\left|
1\right\rangle _{t}$, $\left| 1\right\rangle _{c}\left| 1\right\rangle
_{t}\rightarrow \left| 1\right\rangle _{c}\left| 0\right\rangle _{t}$. Very
recently, the first demonstrations of these linear optics based CNOT\ gates
have been demonstrated \cite{fran03,obrien03,gasparoni04,obrien04,pan04a} to
generate entanglement. The same CNOT operation can also be used to
distinguish the four Bell states \cite{Bruss}: 
\begin{eqnarray}
\left| \phi ^{\pm }\right\rangle _{1,2} &=&\frac{1}{\sqrt{2}}\left( \left|
0\right\rangle _{1}\left| 0\right\rangle _{2}\pm \left| 1\right\rangle
_{1}\left| 1\right\rangle _{2}\right)  \nonumber \\
\left| \psi ^{\pm }\right\rangle _{1,2} &=&\frac{1}{\sqrt{2}}\left( \left|
0\right\rangle _{1}\left| 1\right\rangle _{2}\pm \left| 1\right\rangle
_{1}\left| 0\right\rangle _{2}\right) ,
\end{eqnarray}%
where the subscripts 1 and 2 label different qubits.

In the present work we explicitly realized the controlled disentangling
operation of the CNOT gate for input Bell states from pulsed laser sources.
Our implementation of this gate utilizes the CNOT gate proposed by Pittman
et al. \cite{fran01a} and experimentally demonstrated by Gasparoni et al. %
\cite{gasparoni04}. In particular this gate requires an entangled ancillary
photon-pair\ which in principle gives access to classical feedforward of the
output states as demanded for quantum information tasks \cite{knill01}. This
is in contrast to other destructive linear optics gates \cite%
{fran03,obrien03} which necessarily have to destroy their output states.
Through the usage of in polarization entangled input pairs our experiment
entirely differs from other previous teleportation experiments \cite{pan04a}
where the CNOT was applied to independent qubits, as well as for quantum
process tomography experiments \cite{obrien04} which were constraint to
differently polarized input states of continuous laser sources. Our CNOT
gate is probabilistic with an ideal sucessrate of $\frac{1}{4}$ when
monitoring all four combinations of outputs for the ancilla pair. In the
present version only fixed linear polarizers were used and thus decreasing
the actual success rate to $\frac{1}{16}$. When a Bell state enters a CNOT
gate (Fig. 1) in modes 1 and 2, the state transforms as:%
\begin{eqnarray}
\left| \phi ^{\pm }\right\rangle _{1,2} &\longrightarrow &\left| \pm
\right\rangle _{a}\left| 0\right\rangle _{b}  \nonumber \\
\left| \psi ^{\pm }\right\rangle _{1,2} &\longrightarrow &\left| \pm
\right\rangle _{a}\left| 1\right\rangle _{b},  \label{cpi}
\end{eqnarray}%
where $\left| \pm \right\rangle =\frac{1}{\sqrt{2}}\left( \left|
0\right\rangle \pm \left| 1\right\rangle \right) $ represents the
complementary basis, 1 and 2 label the two input modes, while \emph{a} and 
\emph{b} the two output modes. In the experiment we used the polarization of
photons to represent the qubits with $\left| 0\right\rangle $ denoting the
horizontal polarization state $\left| H\right\rangle $ and $\left|
1\right\rangle $ the vertical polarization state $\left| V\right\rangle $.
In Figure 1, the two input photons and the two photons from an ancilla pair
in the entangled state $\left| \phi ^{+}\right\rangle _{3,4}$ are
superimposed at two polarizing beamsplitters (PBSs). The PBS is an optical
device that transmits horizontally-polarized photons and reflects
vertically-polarized photons. In this experiment, the PBS implements a
two-qubit parity check: if two photons enter the PBS from the two different
input ports, then they must have the same logical value, $\left|
0\right\rangle \left| 0\right\rangle $ or $\left| 1\right\rangle \left|
1\right\rangle $, in order to pass to the two different output ports.

The lower PBS, acting in the $\left| \pm \right\rangle $ basis (indicated by
the circle within the square) performs the logical function of a destructive
CNOT gate, where one of the input qubits is destroyed in the lower detector
(mode \emph{d}) \cite{gasparoni04}. This destruction of one of the input
qubits is compensated by first encoding the value of that qubit onto two
output qubits. This happens by utilizing the parity check at the upper
polarizing beamsplitter and exploiting the entanglement between the modes 3
and 4. Thus conditioned on the successful detection of one and only one
photon in the output mode \emph{c} in the state $\left| +\right\rangle _{c}$
an arbitrary input state in mode $1$ undergoes the following encoding
transformation: $\alpha |0\rangle _{1}+\beta |1\rangle _{1}\rightarrow
\alpha \left| 0\right\rangle _{a}\left| 0\right\rangle _{4}+\beta \left|
1\right\rangle _{a}\left| 1\right\rangle _{4}.$ One of the encoded qubits is
issued as the input in mode 4 for the destructive CNOT gate, while the
remaining copy serves as one of the required logical outputs in mode a. \
The scheme works in those cases where one and only one photon is found in
each of the modes \emph{a},\emph{\ b},\emph{\ c}, and \emph{d} with a
theoretical probability of $\frac{1}{4}$ when the two possible outcomes for
each photon can be distinguished. Therefore, all of the failures in this
CNOT\ gate corresponds to the emission of more than one photon into the same
spatial output mode, which can be arbitrarily reduced by using the quantum
Zeno effect \cite{zeno04}. In the case of photons, where the emission of two
photons into the same output mode must be suppressed, this quantum Zeno
effect can be implemented by utilizing photon-atom interactions, i.e. by an
optical fibre, whose core is containing a single atom. For our
proof-of-principle demonstration, where we used polarizers in the output
beams, the success rate is reduced to $\frac{1}{16}$.

An ultraviolet laser pulse with a wavelength of 395nm passes through a
nonlinear $\beta -$barium borate (BBO) crystal \cite{kwiat95a}, reflects at
a moveable delay mirror, and passes through the crystal for a second time
(Fig. 2). Such an alignment allows the emission of polarization-entangled
photon pairs into the forward pair of modes 3 and 4, as well as into the
backward pair of modes 1 and 2. To counter the effect of birefringence from
the BBO crystal, compensators are used in each mode. These compensators,
composed of a half-wave plate (HWP) performing a 90%
${{}^\circ}$
rotation and an additional BBO crystal, erase the longitudinal and
transversal walk-off of the down-converted photons. Final HWPs, one for each
photon pair, and the tilt of the compensation crystals allow for the
production of any of the four Bell states. The modes of the forward emitted
pairs 1 and 2 and the modes of the backward emitted pairs 3 and 4 are
coherently combined at polarizing beamsplitters (PBSs) by adjusting the
position of the delay mirror. The indistinguishability between the
overlapping photons is improved by introducing narrow bandwidth (3nm)
spectral filters at the outputs of the PBSs and by monitoring the outgoing
photons by single-mode fiber-coupled detectors (Fig. 3). After each PBS a
quarter-wave plate (QWP) compensates birefringence effects in the PBSs.

The destructive CNOT gate is realized by rotating the input states along
modes 2 and 4 by 45$%
{{}^\circ}%
$ through HWPs. Additional HWPs can be used to identify the outcomes
according to the\ original proposal \cite{fran01a}. These HWPs in the output
modes are not necessary if the measurements are made directly in the
complementary $\left| \pm \right\rangle $ basis.

For the Bell state analysis, two Bell states have to be prepared in
different modes and at the same time. These pairs, the input Bell state and
the ancilla pair, have to overlap at the two PBSs. The HWP in mode 4
transforms the ancilla Bell state to $\left| \phi ^{+}\right\rangle
_{3,4}^{rot.}=\frac{1}{\sqrt{2}}(\left| 0\right\rangle _{3}\left|
+\right\rangle _{4}-\left| 1\right\rangle _{3}\left| -\right\rangle _{4})$,
while the HWP in mode 2 transforms each of the individually prepared input
Bell states to 
\begin{eqnarray}
\left| \phi ^{\pm }\right\rangle _{1,2}^{rot.} &=&\frac{1}{\sqrt{2}}(\left|
0\right\rangle _{1}\left| +\right\rangle _{2}\mp \left| 1\right\rangle
_{1}\left| -\right\rangle _{2})  \nonumber \\
\left| \psi ^{\pm }\right\rangle _{1,2}^{rot.} &=&\frac{1}{\sqrt{2}}(-\left|
0\right\rangle _{1}\left| -\right\rangle _{2}\pm \left| 1\right\rangle
_{1}\left| +\right\rangle _{2}).
\end{eqnarray}%
Recalling the action of the each PBS as parity check the four different
input states result in:%
\begin{eqnarray}
\left| \phi ^{\pm }\right\rangle _{1,2}^{rot.}\left| \phi ^{+}\right\rangle
_{3,4}^{rot.} &\longrightarrow &\left| \phi ^{\pm }\right\rangle
_{a,c}\left| \phi ^{+}\right\rangle _{b,d}  \nonumber \\
\left| \psi ^{\pm }\right\rangle _{1,2}^{rot.}\left| \phi ^{+}\right\rangle
_{3,4}^{rot.} &\longrightarrow &\left| \phi ^{\pm }\right\rangle
_{a,c}\left| \phi ^{-}\right\rangle _{b,d},
\end{eqnarray}%
if one photon is detected in each of the four outputs. Projective
measurements in modes \emph{c} and \emph{d} onto the state $\left|
+\right\rangle $ complete the action of the CNOT and transforms each input
Bell state to a different output as follows: $\left| \phi ^{\pm
}\right\rangle _{1,2}\longrightarrow \left| \pm \right\rangle _{a}\left|
+\right\rangle _{b}$ and $\left| \psi ^{\pm }\right\rangle
_{1,2}\longrightarrow \left| \pm \right\rangle _{a}\left| -\right\rangle
_{b} $. Note the similarity between this transformation and Eq. \ref{cpi}.

Each of the four input Bell states were prepared by rotating the HWP and
tilting the compensating BBO in mode 1. The quality of the input state was
quantified by correlation measurements in the $\left| 0/1\right\rangle $ and 
$\left| \pm \right\rangle $ bases, yielding an average fidelity of \ 96\%
and 94\%, respectively. Roughly the same visibilities were obtained for the
ancilla pair when adjusting a $\left| \phi ^{+}\right\rangle _{3,4}$ state
by using the HWP and BBO crystal in mode 3. While the projections onto $%
\left| +\right\rangle $ at the output modes \emph{c} and \emph{d} were kept
fixed, the orientations of the polarizers in the output modes \emph{a} and 
\emph{b} were changed $\left| +\right\rangle _{a}\left| +\right\rangle _{b}$%
, $\left| -\right\rangle _{a}\left| +\right\rangle _{b}$, $\left|
+\right\rangle _{a}\left| -\right\rangle _{b}$, $\left| -\right\rangle
_{a}\left| -\right\rangle _{b}$. Each measurement was made for 1800 s and
yielded a maximum of about 450 four-fold coincidences. \ In contrast to
previous experiments \cite{gasparoni04}, cases where due to the spontaneous
nature of the down-conversion process two pairs are emitted into the same
pair of modes, 1 and 2 or 3 and 4, can ideally never result in a four-fold
coincidence. These\ photons must have orthogonal polarization to be split up
at each PBS to contribute to a four-photon detection after the
beamsplitters. In the present scheme this contribution is suppressed by
destructive quantum interference in the HWP rotation used for preparing the
input Bell states. To explain this in more detail, consider the case that
the source emits for example a $\left| \phi ^{+}\right\rangle _{1,2}$ Bell
state into the modes 1 and 2. When the HWP\ rotates the polarization by $%
\theta $ transforms the Bell state to $\cos (\theta )\left| \phi
^{+}\right\rangle _{1,2}+\sin (\theta )\left| \psi ^{-}\right\rangle _{1,2}$%
, the four-photon contribution after the PBS evolves like $\cos \left(
2\theta \right) \left| 0\right\rangle _{a}\left| 0\right\rangle _{b}\left|
1\right\rangle _{c}\left| 1\right\rangle _{d}$ and thus does not contribute
at our rotation of 45$%
{{}^\circ}%
$. At this specific angle any possible four-photon state, emitted into the
two modes, is excisiting of two photons with the same polarization within
one mode. Thus, these two photons will never be split up at the PBS and
therefore never contribute to a four-fold coincidence detection.

The count rates (Fig. 4) of all 16 possible combinations clearly confirm the
successful implementation of the Bell-state analyzer. The fidelity of the
gate operation $F=N_{correct}/N_{total}$ can be obtained directly\ from the
numbers of the correct measurement results, $N_{correct},$ divided by the
sum of all counts, $N_{total}=N_{correct}+N_{incorrect}.$ The achieved
fidelities of each Bell state analysis are $F_{\phi ^{+}}=(0.75\pm 0.05)$,\ $%
F_{\phi ^{-}}=(0.79\pm 0.05)$, $F_{\psi ^{+}}=(0.79\pm 0.05)$, $F_{\psi
^{-}}=(0.75\pm 0.05)$ without substracting background due to the curious
quantum interference effect. This is a remarkable improvement over former
experiments \cite{gasparoni04}, where the substraction of the signal from
the emission of two pairs into the same pair of modes was required to
achieve high fidelity results. However, minor incorrect outcomes in our
experiment originate mainly from imperfections in the PBSs where photons are
guided incorrectly and from the obtained visibilities when overlapping the
photons at the PBSs (Fig. 3).

For confirmation of the generality of our Bell-state analyzer the input
state superposition $\left| \Psi \right\rangle _{1,2}=\left| \phi
^{+}\right\rangle _{1,2}+i\left| \phi ^{-}\right\rangle _{1,2}$ was created
by tilting the compensating BBO crystal to a position, where the visibility
in the $\left| \pm \right\rangle $ basis was almost 0. The output settings
of the polarizer result in the highest count rates for a $\left| \phi
^{+}\right\rangle $ and a $\left| \phi ^{-}\right\rangle $ setting, 170 and
104 counts, respectively, while the settings for $\left| \psi ^{\pm
}\right\rangle $ input state just measured noise counts.\ This confirms that
the analyzer functions as expected, detecting both the $\left| \phi
^{+}\right\rangle $ and $\left| \phi ^{-}\right\rangle $ components from the
input state.

The experiment reported here explicitly demonstrates for the first time the
action of the CNOT gate on input Bell states, where all Bell states were
identified with an average fidelity of $0.77\pm 0.05$ without any
substraction of noise. Additionally, the experiment shows how quantum
interference effects allow to suppress the spurious signal responsible for
low raw-data fidelities in previous work.

The authors thank K. Resch for discussions. This work is supported by the
Austrian FWF, project F1506 and by the European Commission, project RAMBOQ.

Figure 1. The scheme for the Bell state analyzer (BSA) based on a
nondestructive CNOT gate. The combination of two polarizing beamsplitters
(PBS), one acting in the\ $\left| H/V\right\rangle $\ ($\left|
0/1\right\rangle $)\textbf{\ }basis and the other acting in the complementary%
\textbf{\ }$\left| \pm \right\rangle =\frac{1}{\sqrt{2}}\left( \left|
0\right\rangle \pm \left| 1\right\rangle \right) $\textbf{\ }basis (denoted
by a circle within the square), comprises a non-destructive CNOT gate. Input
Bell states propagate along the modes 1 \& 2, while the photons of the
entangled ancilla pair\textbf{\ }$\left| \phi ^{+}\right\rangle _{3,4}$
propagate along mode 3 \& 4. When the input Bell state overlaps with the
ancilla pair, and one detects one and only one photon in each of modes\emph{%
\ c }\&\emph{\ d}, the analyzer produces for each Bell state a different
product output state which can be easily identified.

Figure 2. The experimental setup. A UV laser pulse at 395nm makes two passes
through a type-II down conversion crystal (BBO) and emits
polarization-entangled photon pairs into the modes 1 \& 2 and 3 \& 4. In
each mode compensators (comp), a combination of a half-wave plate (HWP) and
a second BBO crystal, are used to counter walk-off effects. Final HWPs, one
for each photon pair, and the tilt of the compensating BBO allow for the
production of any of the four Bell-states. The photon pairs in modes 1 \& 2
were prepared in any desired input state $\left| \phi ^{+}\right\rangle
_{1,2}$ , $\left| \phi ^{-}\right\rangle _{1,2}$ , $\left| \psi
^{+}\right\rangle _{1,2}$ and $\left| \psi ^{-}\right\rangle _{1,2}$,\textbf{%
\ }while the ancilla pairs in mode 3 \& 4 were generated in the\textbf{\ }$%
\left| \phi ^{+}\right\rangle _{3,4}$ state. After each PBS a
quarter-waveplate (not shown) was placed into the output modes\emph{\ c }\&%
\emph{\ d }at\textbf{\ }$90\mathbf{%
{{}^\circ}%
}$ to compensate birefringence effects in both PBSs.

Figure 3. ''(Color online)'' Visibilities of the two-photon interference
fringes based on the twofold coincidence measurements after a 45$%
{{}^\circ}%
$ polarizer at the output modes\textbf{\ }\emph{a }\&\textbf{\ }\emph{b}%
\textbf{, }\emph{c }\&\emph{\ d}\textbf{, }\emph{a }\&\emph{\ d}\textbf{\ }%
and\textbf{\ }\emph{c }\&\emph{\ b}\textbf{.} The visibility of the
two-photon interference is shown as a function of the mirror position.

Figure 4. ''(Color online)'' Fourfold coincidences for all possible
combinations of the inputs and outputs are shown. \ Each of the four
different Bell states is transformed into a distinguishable separable state%
\textbf{\ }$\left| \phi ^{\pm }\right\rangle _{1,2}\longrightarrow \left|
\pm \right\rangle _{a}\left| +\right\rangle _{b}$, $\left| \psi ^{\pm
}\right\rangle _{1,2}\longrightarrow \left| \pm \right\rangle _{a}\left|
-\right\rangle _{b}$. Each input state was measured for 1800 seconds at each
of the four different polarizer settings. The fidelity is, on average,%
\textbf{\ }$\left( F=0.77\pm 0.05\right) $.

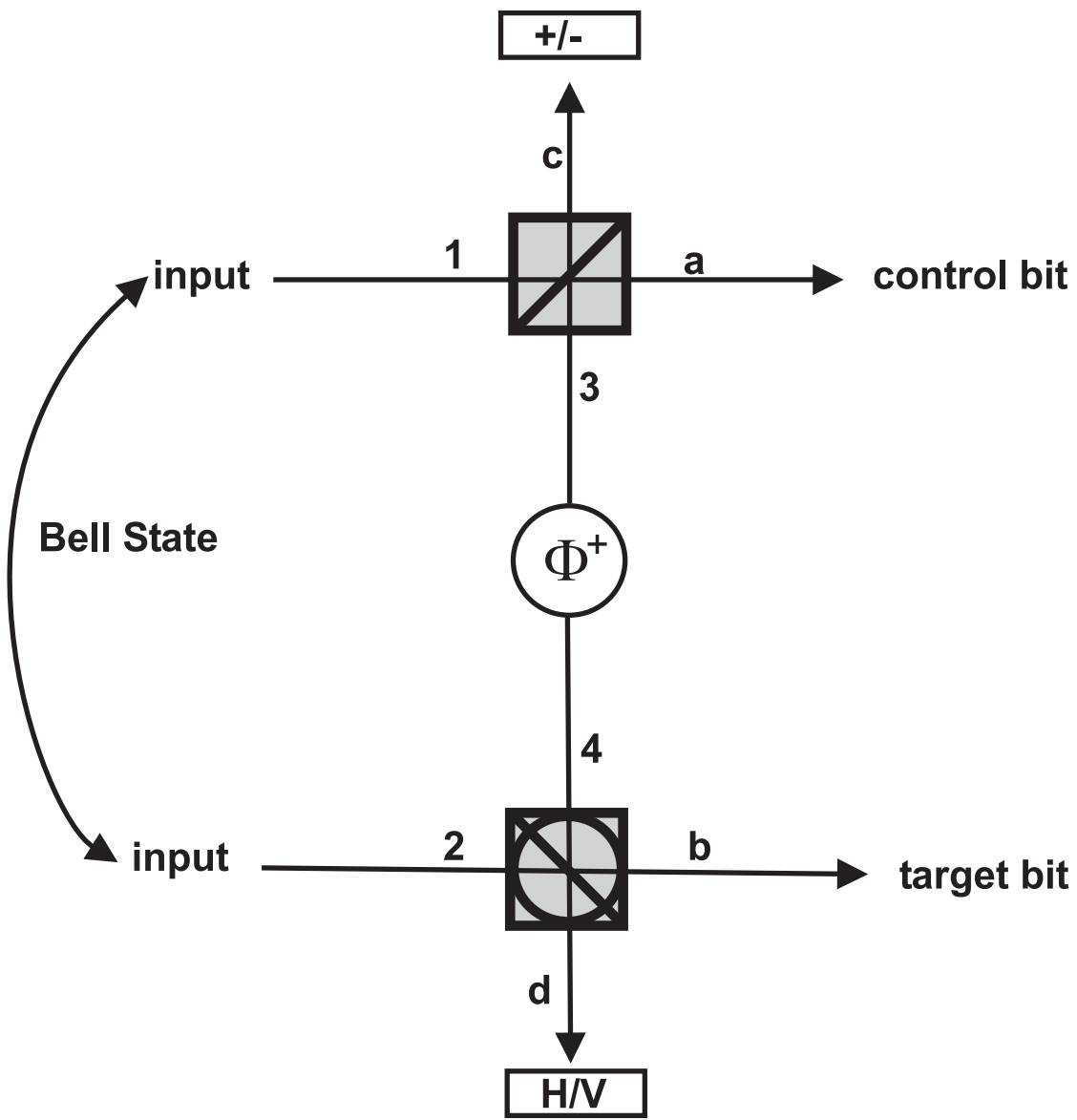
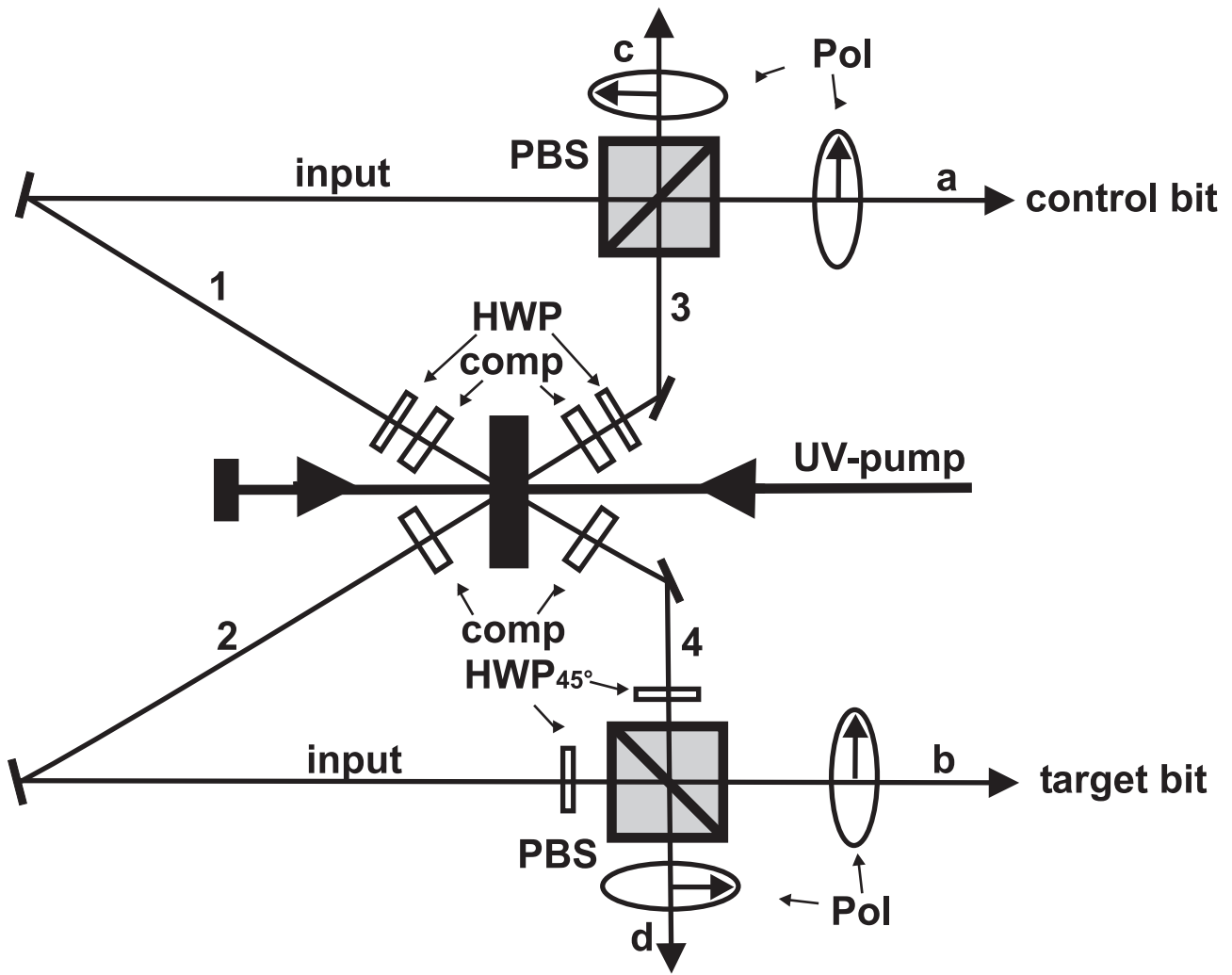
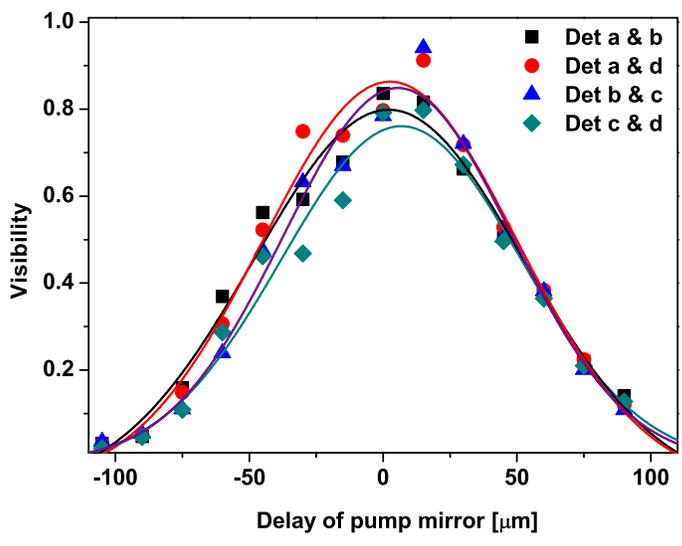
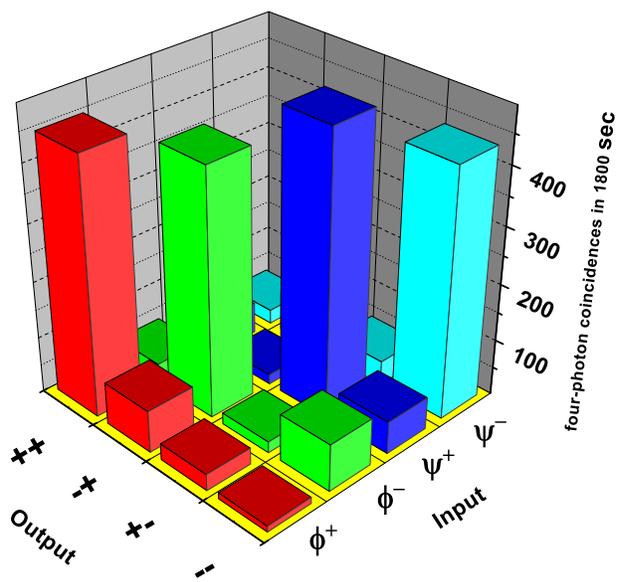
\end{document}